\documentclass[amssymb,prb,twocolumn,showpacs]{revtex4}
\usepackage{epsfig}
\usepackage{dcolumn}
\usepackage{amsmath}
\begin{document}

\title{Magnetoasymmetric current fluctuations of single-electron tunneling}
\author{David S\'anchez}
\affiliation{Departament de F\'{\i}sica,
Universitat de les Illes Balears, E-07122 Palma de Mallorca, Spain}
\date{\today}

\begin{abstract}
We determine the shot noise asymmetry of a quantum dot under reversal
of an external magnetic field. The dot is coupled to edge states
which invert their chirality when the field is reversed, leading
to a magnetoasymmetric electrochemical potential in the nanostructure.
Surprisingly, we find an exact relation between the magnetoasymmetries
corresponding to the nonlinear conductance and the shot noise to leading
order in the applied bias, implying a higher-order fluctuation-dissipation
relationship. Our calculations also show a magnetoasymmetry of
the full probability distribution of the transferred charge.
\end{abstract}

\pacs{73.23.-b, 73.50.Fq, 73.63.Kv}
\maketitle

\section{Introduction}
Recently, it has been theoretically demonstrated \cite{san04,spi04,but05,pol06,mart06,and06}
and experimentally verified \cite{rik05,wei06,mar06,let06,zum06,ang07,har08}
that the Onsager-Casimir reciprocity
relations \cite{ons,cas} cannot be, in general, extended to mesoscopic transport
far from equilibrium. Departures of microreversibility at
equilibrium have been related to the interaction
of the nanostructure with an environment driven out of
equilibrium.\cite{san08} In both cases, it is found that the
current flowing through the system
is not invariant under reversal of the external
magnetic field, leading to the observation of
{\em magnetoasymmetries} solely due to the asymmetric
properties of the internal electrostatic potential.
Thus, the effect is induced purely by electron
interactions.

Previous literature has discussed the size of the
magnetoasymmetries for the electric conductance.
Here, we are concerned with the {\em shot noise},
which is known to offer a complementary and, quite often,
unique tool to probe electronic transport
in quantum correlated nanostructures.\cite{bla00}
Shot noise has been shown to reveal
electronic entanglement detection in mesoscopic interferometers,\cite{los00}
dynamical spin blockade in dots attached to ferromagnetic leads,\cite{cot04}
quantum shuttling in nanoelectromechanical systems,\cite{nov04}
nonequilibrium lifetime broadenings in cotunneling currents,\cite{uts06}
and quantum coherent coupling in double quantum dots,\cite{kie07} 
just to mention a few.

It is worth noting that the linear conductance and
the equilibrium noise are related each other via
the linear fluctuation-dissipation theorem,
which is a general statement 
about the response of a system near equilibrium
and its dynamical fluctuations induced by random forces.\cite{mar08}
For mesoscopic conductors the theorem is expressed as the Johnson-Nyquist
formula between equilibrium current fluctuations
and the linear conductance.\cite{bla00}
Both the linear conductance and the current fluctuations at equilibrium
obey reciprocity relations.\cite{but86} At nonequilibrium, however,
Onsager's microreversibility is generally not fulfilled.
Notably, our calculations explicitly show an {\em exact} relation
between both magnetoasymmetries corresponding to the noise susceptibility
and the nonlinear conductance
to leading order in the applied voltage bias, implying a higher-order
fluctuation-dissipation relationship.
Recent works \cite{foe08,sai08} find
similar fluctuation relations.

\section{The system}

\begin{figure}
\centerline{
\epsfig{file=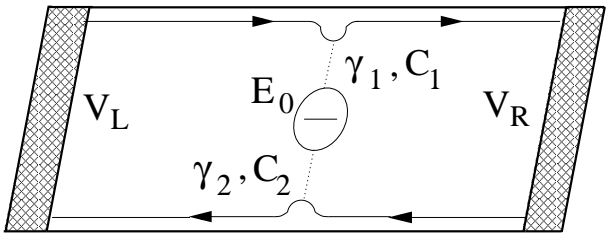,angle=0,width=0.45\textwidth,clip}
}
\caption{Sketch of a quantum dot coupled to chiral edge states
via tunnel and capacitive couplings. The sample is driven
out of equilibrium with an external bias $V_L=-V_R=V/2$.}
\label{figsample}
\end{figure}

We show our result considering a simple but paradigmatic
mesoscopic system---a quantum dot in the Coulomb blockade
regime for which the charge is quantized
and transport is blocked at low temperature unless
charging energy is supplied by external voltage.\cite{kou97,bee91}
We study a dot coupled to two chiral states \cite{for94,kir94}
(filling factor $\nu=1$)
propagating along the opposite edges of a quantum Hall conductor,
as shown in Fig. \ref{figsample}.
For positive magnetic fields $B>0$ carriers in the
upper (lower) edge state move from
the left (right) terminal to the right (left) terminal.
The current flow is reversed for $B<0$.
Coupling between the dot and the edge states takes place
via tunnel couplings $\hbar\gamma_1$ and $\hbar\gamma_2$
and capacitive couplings $C_1$ and $C_2$.

We consider a single energy
level $E_0$ which can be externally tuned with
a gate voltage. ($E_0$ is a kinetic energy
invariant under $B$ reversal). The dot
lies deep in the Coulomb-blockade regime for which 
$E_0+e^2/C$ (with $C=C_1+C_2$)
is well above the electrochemical potentials
$\mu_\alpha=E_F+eV_\alpha$ ($\alpha=L,R$)
of both left ($L$) and right ($R$) leads,
$E_F$ being the common Fermi energy.
We assume that each edge state
is in equilibrium with its corresponding
injecting reservoir. Therefore, they act
effectively as massive electrodes with
well defined electrochemical potentials.
Without loss of generality, we take
$V_L=-V_R=V/2$ with $V$ the applied bias voltage.

For temperatures $k_B T$ and voltages low enough, $k_B T,eV\ll e^2/C$
(but $k_B T$ sufficiently high to neglect Kondo correlations),
the charging energy $e^2/C$ is the dominant
energy scale of the problem and double occupation
in the dot is negligible.
In the master equation approach, the charging
state of the dot is given by an integer number of
electrons, $n$. We assume that quantum coherence
between states with differing $n$ is lost,
as occurs in the pure Coulomb blockade regime.
For definiteness, we consider spin polarized carriers,
although the model can be easily extended to the
spinful case. Thus, the dot can be either
in the empty state ($n=0$) or in the
occupied state ($n=1$) with instantaneous occupation
probability $p_n(t)$ at time $t$.
The main transport mechanism is via single-electron tunneling
($\gamma_1,\gamma_2\ll k_B T/\hbar$)
and we can write quantum rate equations for $p_0(t)$
and $p_1(t)$,\cite{her93,han93,kor94}

\begin{equation}\label{rate}
\left( \begin{array}{c}
\dot{p}_0 \\
\dot{p}_1\end{array} \right)=
\left( \begin{array}{cc}
-\Gamma^+ & \Gamma^-\\
\Gamma^+ & - \Gamma^-\end{array} \right) 
\left( \begin{array}{c}
p_0 \\
p_1\end{array} \right)
\,.
\end{equation}

In a more compact form, one has $\dot{\vec{p}}=\mathbf{M} \vec{p}$, where $\vec{p}$
is the vector with components $\vec{p}=(p_0\;p_1)^T$
and the $2\times 2$ matrix $\mathbf{M}$ can be obtained from
Eq.~(\ref{rate}). We note that the columns of $\mathbf{M}$ add to zero to fulfill
the condition $p_0+p_1=1$ at every instant of $t$.
The elements of $\mathbf{M}$ represent
transition rates, $\Gamma^\epsilon$, to tunnel on
($\epsilon=+1$) and
off the dot ($\epsilon=-1$). The total rates are given by
$\Gamma^\epsilon=\sum_{\alpha}\Gamma_\alpha^\epsilon$.
For $B>0$ we find (see Fig. \ref{figsample})
$\Gamma_L^+=\gamma_1 f(+B,V_L)$,
$\Gamma_R^+=\gamma_2 f(+B,V_R)$,
$\Gamma_L^-=\gamma_1 [1-f(+B,V_L)]$
and $\Gamma_R^-=\gamma_2 [1-f(+B,V_R)]$
where the occupation factors are given by
$f(B,V_\alpha)=1/(1+\exp{(\mu_d(B)-eV_\alpha)/k_B T})$
for $E_F=0$. The electrochemical potential in the dot,
$\mu_d(B)$, is self-consistently found from the electrostatic
configuration, which depends on the $B$ direction.\cite{san05}
For $B>0$, one has $\mu_d(+B)=E_0-eC_1V_L/2C-eC_2V_R/2C$.

For $B<0$ the chirality of the edge states is inverted
and the nonequilibrium ($V_L\neq V_R$)
polarization charge changes accordingly.
Injected electrons in the upper (lower) edge state
are now predominantly
emitted from lead $R$ ($L$).
Therefore, the rates now read
$\Gamma_L^+=\gamma_2 f(-B,V_L)$,
$\Gamma_R^+=\gamma_1 f(-B,V_R)$,
$\Gamma_L^-=\gamma_2 [1-f(-B,V_L)]$
and $\Gamma_R^-=\gamma_1 [1-f(-B,V_R)]$,
where $\mu_d(-B)=E_0-eC_2V_L/2C-eC_1V_R/2C$.
We stress that the rates depend, quite generally, on the $B$ direction
and differ for $C_1\neq C_2$.
As demonstrated in Ref. \onlinecite{san05},
the magnetoasymmetry of the polarization charge
leads to a magnetoasymmetric addition energy
of the quantum dot and, as a result,
the current average is an uneven function of $B$.
We can thus ask ourselves whether
the current fluctuations also
exhibit this effect.

Current fluctuations are characterized by the
power spectrum, $S_{\alpha\beta}(\omega)$,
of the current--current
correlation function,\cite{bla00}
\begin{equation}\label{eq_sn}
S_{\alpha\beta}(\omega)=2\int e^{i\omega t}
[\langle I_\alpha(t) I_\beta(0)\rangle-
\langle I_\alpha\rangle\langle I_\beta\rangle]\,dt
\,,
\end{equation}
where $\langle I_\alpha\rangle$ is the averaged current
across the $\alpha$ junction. Its stationary value
can be found from the steady state solution of Eq.~(\ref{rate}),
which reads $\bar{p}_0=\Gamma^-/\Gamma$ and
$\bar{p}_1=\Gamma^+/\Gamma$. As a consequence,
\begin{equation}\label{eq_cur}
\langle I_\alpha\rangle=e[\Gamma_\alpha^+\bar{p}_0
-\Gamma_\alpha^-\bar{p}_1]=e
\frac{\Gamma_\alpha^+\Gamma^--\Gamma_\alpha^-\Gamma^+}{\Gamma}\,,
\end{equation}
where $\Gamma=\Gamma^++\Gamma^-$.

The correlator $\langle I_\alpha(t) I_\beta(0)\rangle$
in Eq.~(\ref{eq_sn}) can be obtained from the conditional
probability $p_{n,n'}(t)$ that the state $n$ is occupied
at time $t>0$ when the dot was in the state $n'$ at $t=0$.
Within our scheme the quantum regression theorem holds,
and $p_{n,n'}(t)$ obeys the same equations as $p_n(t)$.
Hence, the eigenvalues of $\mathbf{M}$ completely determine the
dynamical behavior of $p$. This can be seen from the
noise expression
$S_{\alpha\beta}=S_\alpha^{\rm Sch} \delta_{\alpha\beta}
+ S_{\alpha\beta}^c$,
with $S_\alpha^{\rm Sch}$
the Schottky noise produced by correlated tunneling
through a single junction and
\begin{equation}\label{eq_sn3}
S_{\alpha\beta}^c(\omega)=\sum_{\epsilon,\epsilon'}
\epsilon\epsilon'\bar{p}_\frac{1-\epsilon}{2}
[\Gamma_\alpha^{\epsilon'} \Gamma_\beta^{\epsilon} 
+ \Gamma_\beta^{\epsilon'} \Gamma_\alpha^{\epsilon}]
G_{\frac{1-\epsilon'}{2},\epsilon}(\omega)
\,,
\end{equation}
where the ``Green function'' matrix for Eq.~(\ref{rate}) reads,
\begin{equation}\label{eq_g}
\mathbf{G(\omega)}=-{\rm Re} (i\omega + \mathbf{M})^{-1}\,.
\end{equation}
Care must be taken with the $\omega=0$ limit since
$\mathbf{M}$ is singular.
The matrix $\mathbf{M}$ has two eigenvalues, namely,
$\lambda_1=0$ with an eigenvector given by the
stationary solution $\vec{p}=(\Gamma^-/\Gamma\; \Gamma^+/\Gamma)^T$
and $\lambda_2=-\Gamma$, which describes a charge excitation in the system.
Therefore, we can now use in Eq.~(\ref{eq_g})
the spectral decomposition
$\mathbf{M}=\sum_{i=1,2} \lambda_i \mathbf{U} \mathbf{E}_i \mathbf{U}^{-1}$,
where $\mathbf{E}_i$ is a matrix with the element $(i,i)$ equal to 1
and the other elements are zeroes,
the $i$-th column of $\mathbf{U}$ being the $i$-th eigenvector of $\mathbf{M}$.
The $\lambda_1$ contribution cancels out the term
$\langle I_\alpha\rangle\langle I_\beta\rangle$. As a result,
$\mathbf{G}(\omega)=-\lambda_2 \mathbf{U} \mathbf{E}_2 \mathbf{U}^{-1}/(\omega^2+\lambda_2^2)$
and we find the shot noise $S=-S_{LR}$ at $\omega=0$,
\begin{equation}\label{eq_noise0}
\frac{S}{2e^2}=\frac{(\Gamma_L^+\Gamma_R^-+\Gamma_L^-\Gamma_R^+)\Gamma^2
-2(\Gamma_L^+\Gamma_R^--\Gamma_L^-\Gamma_R^+)^2}{\Gamma^3}
\,.
\end{equation}
For $B>0$ and large bias such that $eV\gg E_0,k_B T$,
we have $\Gamma_L^+\approx \gamma_1$, $\Gamma_L^-\approx 0$,
$\Gamma_R^+\approx 0$ and $\Gamma_R^-\approx \gamma_2$.
Substituting these values in Eq.~(\ref{eq_noise0})
we recover the double barrier case for noninteracting electrons,
$S=2e^2 \gamma_1 \gamma_2
(\gamma_1^2+ \gamma_2^2)/\gamma^3$,
where $\gamma=\gamma_1+\gamma_2$.
\begin{figure}
\centerline{
\epsfig{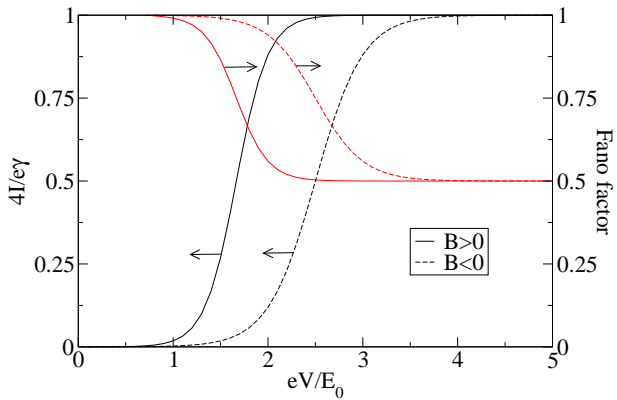}
}
\caption{(Color online) Current (black curves) and Fano factor (red curves)
as a function of bias voltage for $\gamma_1=\gamma_2=0.005 E_0/\hbar$,
$k_B T=0.1 E_0$, $C_1=0.6$ and $C_2=0.4$.}
\label{fig1}
\end{figure}

\section{Results}

Figure~\ref{fig1} shows results for the averaged current
$I=\langle I_L\rangle=-\langle I_R\rangle$
and the Fano factor $F=S/2eI$ as a function of $V$.
The current is exponentially suppressed at low $V$,
increases at $V\sim C E_0/eC_1$ and reaches the limit value
$e\gamma_1 \gamma_2 /\gamma$ for large voltages.
When $B$ is reversed, the current now increases at
$V\sim C E_0/eC_2$. Since we choose $C_1>C_2$ the $I(V,-B)$
is shifted to larger voltages compared to $I(V,+B)$.
As a consequence, the differential conductance is generally
$B$-asymmetric.
The Fano factor is Poissonian at small $V$
since transport is dominated by thermal activated tunneling.
For increasing $V$, the noise becomes sub-Poissonian
and reaches saturation for large $V$. The crossover step
from Poissonian noise to sub-Poissonian noise has a width
which depends on $k_B T$. Like the current, the crossover
center shifts to larger voltages when $B$ is reversed,
thus yielding a magnetoasymmetric Fano factor.

\begin{figure}
\centerline{
\epsfig{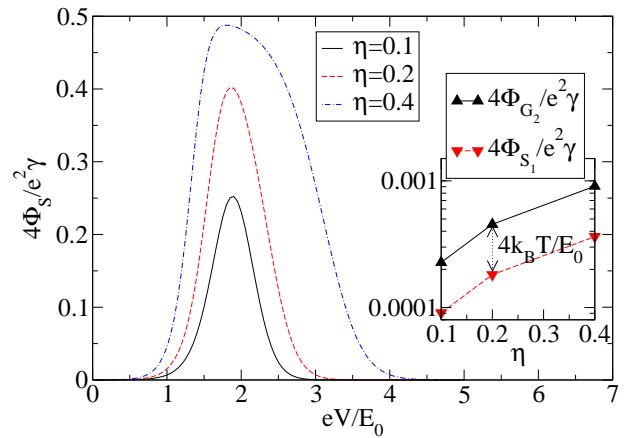}
}
\caption{(Color online) Shot noise magnetoasymmetry for various capacitance asymmetries
$\eta=(C_1-C_2)/(C_1+C_2)$. We take $\gamma_1=\gamma_2=0.005 E_0/\hbar$ and
$k_B T=0.1 E_0$. Inset: Magnetoasymmetries
of the leading-order nonlinear conductance and
nonequilibrium noise at $V=0.03E_0/e$.}
\label{fig2}
\end{figure}

Since the Fano factor depends on both the noise and the
current and these are asymmetric under $B$ reversal,
we plot in Fig.~\ref{fig2} the magnetic field asymmetry
for the noise alone, which we define as $\Phi_S=[S(+B)-S(-B)]/2$.
The asymmetry vanishes for small $V$, fulfilling
the Onsager symmetry. At large $V$
the noise saturation value is independent of the $B$
direction since this limit corresponds to noninteracting
fermions. As a result, the asymmetry vanishes.
The asymmetry becomes maximal for intermediate voltages.
Importantly, the asymmetry increases with the
capacitance asymmetry, $\eta=(C_1-C_2)/C$.
Therefore, the current fluctuations are magnetoasymmetric
only in the case where the electrostatic coupling of the dot
with the edge states leads
to an asymmetric screening of charges.

Furthermore, we consider the case of low voltages,
$eV\ll k_B T$. Then, we can expand $I$ and $S$ in
powers of $V$,
\begin{eqnarray}
I&=&G_1 V+G_2 V^2+\mathcal{O}(V^3)\,, \\
S&=&S_0 + S_1 V+\mathcal{O}(V^2)\,. \label{s1}
\end{eqnarray}
We have checked that both the linear conductance $G_1$ and
the equilibrium noise $S_0$ are even functions of $B$.
They satisfy the {\em linear} fluctuation dissipation
theorem, $S_0=4 k_B T G_1$.
While $G_1$ depends on the equilibrium potential in the dot,
the nonlinear conductance term $G_2$ is, in general, a function
of the screening electrostatic potential.\cite{christen}
This potential need not be $B$-symmetric.\cite{san04} Thus,
the magnetoasymmetry $\Phi_{G_2}=[G_2(B)-G_2(-B)]/2$
acquires a finite value:
\begin{equation}\label{phig2}
\Phi_{G_2}=\frac{e^3}{8}
\frac{\gamma_1 \gamma_2}{\gamma}
\frac{\eta}{(k_B T)^2} {\rm sech}^2\frac{E_0}{2k_BT}\tanh\frac{E_0}{2k_BT}
 \,.
\end{equation}
For high $k_BT$, Eq.~(\ref{phig2}) yields
$\Phi_{G_2}=0$ since thermal fluctuations
are $B$-symmetric. $\Phi_{G_2}$ determines the rate
at which the current magnetoasymmetry increases
with voltage, thus showing that magnetoasymmetries
are a truly nonequilibrium effect.

The nonequilibrium noise
at linear response [i.e., $S_1$ in Eq.~(\ref{s1})] is also $B$-asymmetric.
To leading order in $V$, we find
\begin{equation}\label{eqs1g2}
\Phi_{S_1}=4 k_B T\Phi_{G_2}\,,
\end{equation}
where $\Phi_{S_1}=[S_1(B)-S_1(-B)]/2$.
This is a relevant result of our work.
Remarkably, we obtain the same functional dependence for
the magnetoasymmetries of noise and current
in the leading-order nonlinearities.
We numerically confirm this prediction for
a small value of $V$ (see inset of Fig.~\ref{fig2}).
This result can be related to the {\em nonlinear}
fluctuation-dissipation theorem, which reads $S_1=4 k_B T G_2$. It has been derived
in Ref. \onlinecite{tob07} for mesoscopic conductors
at arbitrary voltages within
the framework of full counting statistics.
Thus, it is shown\cite{tob07} that the nonlinear fluctuation-dissipation theorem holds
even for interacting particles assuming time-reversibility
(no magnetic fields). The nontrivial
difference, however, is that in our theory
microreversibility is broken due to the combined effect
of magnetic fields and interactions \cite{san04} but still
Eq.~(\ref{eqs1g2}) holds. While detailed balance conditions are shown
to hold far from equilibrium in the absence
of magnetic fields,\cite{tob07,andr06,esp07} the same relations can not,
generally, be established when microreversibility
is broken.\cite{foe08} Despite this, we obtain
an unexpected symmetry relation between the
conductance and noise response magnetoasymmetries.

We note in passing that Eq.~(\ref{eqs1g2})
is not a generalized of the fluctuation-dissipation theorem for which
nonlinear fluctuations and the response are
related via a (nontrivial) effective temperature,
as in glassy systems, \cite{kob97} since the prefactor for
both linear and nonlinear theorems is the same.
Our result also differs from more general
fluctuation theorems obeyed by full probability
distributions.\cite{eva93}
\begin{figure}
\centerline{
\epsfig{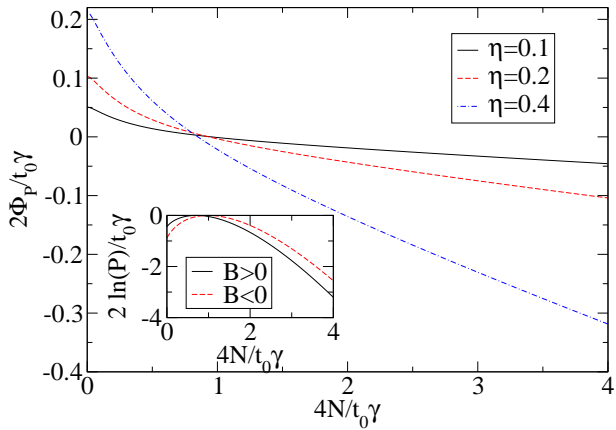}
}
\caption{(Color online) Charge probability distribution magnetoasymmetry
for $\gamma_1=\gamma_2=0.05 E_0/\hbar$,
$k_B T=0.4 E_0$, $eV=-4.5E_0$, and various values of $\eta$.
Inset: $P(N)$ for two field orientations and $\eta=0.4$.}
\label{fig3}
\end{figure}

To analyze higher-order terms one should
consider fluctuations of the screening potential,
which are beyond the scope of a mean-field approximation.
Nevertheless, for classical Coulomb blockade effects
the local potential fully screens the excess charges
and quantum fluctuations are absent. Therefore,
our model system is perfectly suitable for further
extensions. We note that the Fano factor magnetoasymmetry,
$\Phi_{F}=[F(B)-F(-B)]/2$,
is {\em quadratic} in voltage:
\begin{equation}
\Phi_{F}=
\frac{\gamma_1 \gamma_2}{\gamma^2}\eta
\left(\frac{eV}{2k_B T}\right)^2 {\rm sech}^2\frac{E_0}{2k_BT}\tanh\frac{E_0}{2k_BT}
 \,.
\end{equation}

A complete characterization of current fluctuations
is given by the full counting statistics,\cite{lev93} which
yields the entire probability distribution $P(N)$
of the transferred charge during the measurement time $t_0$.
We follow the method of Bagrets and Nazarov \cite{bag03}
to assess the cumulant generating function
$\mathcal{S}(\chi)=-\ln \sum_N P(N) e^{iN\chi}$.
Without loss of generality, we count charges in the $R$ lead.
Therefore, we make the substitutions $\Gamma_R^-\to\Gamma_R^-e^{i\chi}$
and $\Gamma_R^+\to\Gamma_R^+e^{-i\chi}$ in the
off-diagonal elements of $\mathbf{M}$
and $\mathcal{S}(\chi)$ is derived from
$\mathcal{S}(\chi)=t_0\lambda_2(\chi)$.
We calculate $P(N)$ within the saddle-point approximation,
valid in the limit $t_0\to\infty$ \cite{bag03}
and determine the magnetoasymmetry of $P(N)$.
We show $\Phi_P=[P(N,+B)-P(N,-B)]/2$ in
Fig. \ref{fig3}. $|\Phi_P|$ increases for increasing
capacitance asymmetry and vanishes around $N\sim \gamma t_0/4$.
This point corresponds to the mean current $\bar I=e\gamma/4$
for a dot symmetrically coupled ($\gamma_1=\gamma_2$)
in the limit $eV\gg k_BT$.

\section{Conclusions}

To summarize, we have investigated magnetoasymmetric
current fluctuations of a Coulomb-blockaded quantum dot.
It is well established that the linear fluctuation-dissipation theorem
makes an equivalence between linear-response functions to small
perturbations and correlation functions describing fluctuations
due to electric motion. In this work, we have found a similar
fluctuation-dissipation relation that predicts an exact equivalence
between the leading-order rectification and noise magnetoasymmetries,
valid in the presence of external magnetic fields.
Such relation has been very recently shown to derive
from fundamental principles.\cite{foe08}
Moreover, we have shown that the full probability distribution
associated to the flow of charges is, generally, magnetic-field asymmetric.
Since the effect studied here relies purely on interaction,
it should be observable in many other systems exhibiting
strong charging effects.

\section*{Acknowledgements}
I thank M. B\"uttiker, H. F\"orster and R. L\'opez for helpful discussions.
This work was supported by the Spanish MEC Grant No.\ FIS2005-02796
and the ``Ram\'on y Cajal'' program.

\end{document}